# Preventing an Extractive Green Hydrogen Industry: Risks and Benefits of Grid Expansion and Green Hydrogen in and for Kenya

Xi Xi[1, 4]*, Boniface Kinyanjui[2], Daniel M. Kammen[1, 3, 4, 5, 6]

[1] Energy and Resources Group, University of California, Berkeley, CA, USA

[2] Energy and Petroleum Regulatory Authority, Nairobi, Kenya

[3] Goldman School of Public Policy, University of California, Berkeley, CA, USA

[4] Department of Civil and Systems Engineering, Johns Hopkins University, Baltimore, MD, USA

[5] Paul Nitze School of Advanced International Affairs, Johns Hopkins University, Washington, DC USA

[6] The Ralph O'Connor Institute of Sustainable Energy, Johns Hopkins University, USA

ABSTRACT

This study evaluates the role of grid-connected hydrogen electrolyzers in advancing a cost-effective and in particular an equitable green hydrogen industry in Kenya to serve both domestic and international needs and markets. Using a multi-nodal capacity expansion model with county-level spatial resolution, we assess how electrolyzer deployment affects electricity cost, grid flexibility, and carbon intensity under various renewable and demand scenarios. Results show that electrolyzers enable up to 30 percent reduction in levelized cost of electricity (LCOE) and $460 million in cumulative system cost savings by 2050 compared to a business-as-usual scenario. As a flexible demand available to absorb surplus generation, electrolyzers reduce curtailment and support large-scale wind integration while still requiring a diverse mix of renewable electricity. The resulting hydrogen reaches a levelized cost of $3.2/kg by 2050, and its carbon intensity from electricity use falls below one kg $CO_2$e/kg $H_2$, suggesting likely compliance with international certification thresholds. Benefits persist across all demand trajectories, though their scale depends on the pace of wind expansion. Spatial analyses reveal unequal distribution of infrastructure gains, underscoring the need for equity-oriented planning. These findings suggest that grid-integrated hydrogen, if planned in coordination with wind investment, transmission, and equitable infrastructure deployment, can reduce costs, support certification, and promote a more equitable model of hydrogen development. In other words, connecting electrolyzers to the grid will not only make green hydrogen *in* Kenya but also *for* Kenya.

# SYNOPSIS

Green hydrogen can reduce emissions or risk reinforcing development challenges under climate goals. This study shows that grid-integrated hydrogen can lower costs, support renewables, and serve Kenya's development needs.

# 1 Introduction

Green hydrogen – hydrogen produced through electrolysis powered by renewable energy sources – has emerged as a critical technology to facilitate the decarbonization of the global energy system. Because of its ability to mitigate emissions in hard-to-abate sectors such as steel, cement, and heavy transport[1], many countries have prioritized green hydrogen as part of their climate and industrial development agendas. More than ten African countries have published national green hydrogen strategies, joined the Africa Green Hydrogen Alliance, or identified potential green hydrogen projects.[2–4] Despite billions of dollars committed to these initiatives, most financing originates outside the continent, with Europe emerging as a dominant funder and expected beneficiary—positioning African hydrogen exports as a mechanism for meeting European decarbonization targets.[3,5]

The European role and in fact dominance in African green hydrogen ventures raises questions of colonialism, financial leverage, and injustice, particularly given widespread energy poverty. Most African countries still lack universal electricity access.[6] Even those with nominal access, such as Tunisia—where energy import dependence surged from 5 percent in 2010 to 50 percent by 2022—face significant supply insecurity while proposing hydrogen exports.[7] Critics argue that these initiatives risk perpetuating a new form of energy colonialism—where African resources are extracted to meet foreign decarbonization needs, often without commensurate local benefit. These dynamics strongly echo historical patterns of externally driven resource exploitation across the continent.[8]

In September 2023, Kenya unveiled its Green Hydrogen Strategy and Roadmap ("the Kenyan strategy") – a collaborative effort between the Kenyan government and, notably, European partners, joining a growing number of African nations with formal green hydrogen plans.[9]

Among all African countries who have published national green hydrogen strategy, Kenya is the only country whose policy exclusively focuses on domestic use in derivative industries. However, the Kenyan strategy suggests that domestic focus is motivated by the high levelized cost of hydrogen (LCOH) produced in Kenya, predominantly driven by its high electricity cost. The strategy envisions domestic production of green hydrogen derivatives, particularly green methanol and fertilizer, for local and regional markets. With agriculture as a cornerstone of Kenya's economy, such efforts could reduce import dependence and insulate farmers from volatile global fertilizer prices currently negatively affecting the country's agriculture sector.[10] Replacing fossil-based fertilizers with green alternatives would also generate substantial climate benefits.

A domestic strategy without addressing high energy costs can cause harm to Kenya's economy and exacerbate existing cost burden to farmers. Multiple studies have shown that green hydrogen must reach the price of $2/kg to produce competitive derivative products ("cost parity").[11,12] The Kenyan strategy shows that the cheapest LCOH achievable in Kenya is around $4-5/kg, with electricity costs accounting for 50-70 percent of the total.[9] At this price, green hydrogen does not reach cost parity. Achieving competitiveness would require either targeted subsidies or consumption mandates. Government subsidies would add pressure to Kenya's already constrained public finances, while mandates would unjustly shift the financial burden onto farmers—many of whom have contributed little to global emissions but face rising fertilizer prices.[13,14] High electricity cost, the major barrier to a viable green hydrogen industry, also constrain broader national development, as stated by the country's fourth Medium Term Plan (MTP IV).[13]

While standalone green hydrogen projects, such as those proposed in the current Kenyan strategy, face high risks of economic failure and equity concerns, grid-connected electrolyzers offer a more integrated pathway that can reduce electricity costs and improve hydrogen economics. Standalone projects require dedicated renewable energy plants to power hydrogen production, limiting their flexibility and providing few opportunities for cost reduction absent major technological breakthroughs. In contrast, grid-connected electrolyzers function as large, flexible power consumers that can improve grid operations and influence long-term capacity planning.

A growing body of literature has assessed the implications of green hydrogen for power system planning, though most quantitative analyses to date focus on developed-country grids[15–17] or hydrogen use for heating[18] and long-duration energy storage (LDES)[19]. These applications are less relevant for Kenya, where heating demand is low and long-term variability is not a major concern for power planning. However, short-term flexibility remains a critical challenge for the Kenyan grid. Grid-connected hydrogen electrolyzers can provide such flexibility, helping to integrate greater shares of variable renewable energy.[15,20–22] Increasing share of wind generation, which is typically more cost-effective than geothermal for each unit of electricity generated, can lower electricity costs and expand national generation capacity, consistent with the priorities outlined in Kenya's MTP IV.[13] This potential is further supported by Carvallo et al., who demonstrate that demand response strategies can help utilize high wind availability during shoulder hours (i.e. hours between peak and valley of demand) in Kenya.[23]

This study employs a multi-nodal capacity expansion model to evaluate the hypothesis that grid-connected hydrogen electrolyzers can reduce electricity costs by providing operational flexibility and enabling expanded wind deployment. In doing so, the model also assesses whether

Kenya's grid mix can support certification of hydrogen as "green" under international standards. Finally, the model's spatial resolution allows for the examination of how hydrogen infrastructure affects regional equity, illuminating potential distributive justice issues associated with a grid-integrated hydrogen transition.

## 2 Methods and Data

### 2.1 Capacity Expansion Model

Several studies have examined Kenya's grid planning and green hydrogen strategies, none has successfully integrated the two to evaluate the flexibility and cost impacts that electrolyzers may bring. Carvallo et al. (2017) were the first to employ Switch 1.0 in Kenya with county-level resolution, evaluating least cost expansion plans and balancing generation and transmission costs against operational and environmental impacts.[23] Since then, the Switch model and input datasets for Kenya, such as installed capacity and various cost inputs, have changed. Most recently, Kihara et al. employed a soft link approach between OSeMOSYS and FlexTool to achieve least-cost planning for a high renewable grid such as Kenya's at the national model.[24] Their OSeMOSYS model used a single-node framework for long-term planning, while FlexTool, also a linear optimization tool, was linked to capture flexibility constraints critical for integrating non-dispatchable renewables. However, the location of electrolyzers may have significant impact on the transmission grid, revealing differentiated impacts across the country. A single-node model does not capture these effects and is therefore insufficient for the objectives of this study.

Lubello et al. subsequently used OSeMOSYS to analyze potential local applications of green hydrogen in fertilizer and iron and steel industries, considering location-specific renewable resource availability[25]. However, their analysis predetermines one specific location for electrolyzers with a dedicated renewable power plant. Müller et al. applied a geospatial optimization framework for Kenya to identify least-cost locations for a variety of green hydrogen production and utilization technologies.[26] Ishmam et al. designed a multidisciplinary framework for mapping green hydrogen potential in Sub-Saharan Africa, considering land eligibility, renewable energy potential, water supply, and other socio-economic impacts, using a region in Benin as a case study.[27] While these studies contribute valuable insights, none explicitly examine grid-connected hydrogen production. As a result, they do not evaluate the role of electrolyzers in providing grid flexibility, nor do they assess how added generation capacity might benefit the broader electricity system or Kenyan consumers.

To quantify the overall and distributive impacts of grid-connected hydrogen electrolyzers, this study employs the Switch 2.0 power system planning model, an open-source power system planning model widely applied to countries with diverse grid characteristics. Switch has been used across multiple continents—including Africa, Asia, Europe, and the Americas—and in countries at different income levels and stages of power system development.[28] It is a mixed integer linear program that minimizes system cost while meeting various investment, operational, and policy constraints. The core Switch model optimizes over total capital and operational costs of the entire system over the planning period. The objective function can be simplified as:

$$\min C = \sum_{p \in P, g \in G, z \in Z} \frac{1}{(1+r)^{Y_p - Y_0}} \left[ \sum_{g \in G} (InvCost_{g,p,z} + FOMCost_{g,p,z}) + \sum_{t \in T_p} w_{p,t}(VOMCost_{g,t,z} + FuelCost_{g,t,z}) \right],$$

Where:

- $C$:              Total system cost (objective to minimize)
- $p \in P$:        Investment periods
- $g \in G$:        Generation, transmission, and storage projects

- $z \in Z$:                Load zones
- $r$:                      Discount rate
- $Y_p$:                    Year corresponding to period $p$
- $Y_0$:                    Reference year for discounting (start year)
- $t \in T_p$:              Timepoints or time slices within period $p$
- $InvCost_{g,p}$:          Investment cost for project $g$ in period $p$
- $FOMCost_{g,p}$:          Fixed O&M cost for project $g$ in period $p$
- $w_{p,t}$:                Weight of timepoint $t$ in period $p$
- $VOMCost_{g,t}$:          Variable O&M cost for project $g$ in timepoint $t$
- $FuelCost_{g,t}$:         Fuel cost for project $g$ in timepoint $t$

The core model does not include hydrogen electrolyzers. However, it is modular and can be extended with user-defined components. We created a hydrogen electrolyzer module that includes construction and operation of hydrogen electrolyzers and sale of hydrogen. Electrolyzer capacity can be capped in each investment period to reflect potential constraints related to financing, governance, or market demand. Capital and operational costs of electrolyzers are incorporated into the existing cost structure of the core model. Revenue from hydrogen sales is treated as a negative cost in the objective function. This revenue is calculated as the product of hydrogen output and the assumed hydrogen price in each investment period. The modified objective function is:

$$\min C = \sum_{p \in P, g \in G, z \in Z} \frac{1}{(1+r)^{Y_p - Y_0}} \left[ \sum_{g \in G} (InvCost_{g,p,z} + FOMCost_{g,p,z}) + \sum_{t \in T_p} w_{p,t}(VOMCost_{g,t,z} + FuelCost_{g,t,z} - H2Rev_{t,z}) \right],$$

Where:

- $C$:          Total system cost (objective to minimize)
- $p \in P$:    Investment periods
- $g \in G$:    Generation, transmission, storage, and electrolyzer projects
- $z \in Z$:    Load zones
- $r$:          Discount rate

- $Y_p$: Year corresponding to period $p$
- $Y_0$: Reference year for discounting (start year)
- $t \in T_p$: Timepoints or time slices within period $p$
- $InvCost_{g,p,z}$: Investment cost for project $g$ in period $p$ and load zone $z$
- $FOMCost_{g,p,z}$: Fixed O&M cost for project $g$ in period $p$ and load zone $z$
- $w_{p,t}$: Weight of timepoint $t$ in period $p$
- $VOMCost_{g,t,z}$: Variable O&M cost for project $g$ in timepoint $t$ and load zone $z$
- $FuelCost_{g,t,z}$: Fuel cost for project $g$ in timepoint $t$ and load zone $z$
- $H2Rev_{t,z}$: Revenue from hydrogen sales in timepoint $t$ and load zone $z$

An additional customized module was added to control capacity limits for specific technologies during user-defined investment periods. In the Switch model, generation technologies and generation projects are represented as separate sets. Multiple projects may share the same technology type, but project-specific parameters such as capital cost and capacity factor can vary due to locational differences. For instance, geothermal project costs differ substantially by site, while wind and solar projects exhibit site-dependent variability in capacity factors. Kenya's most recent Least Cost Power Development Plan (LCPDP) outlines capacity limits by technology type, but these limits are often not specified at the project or county level.[24] The capacity limit module ensures that modeled investment decisions do not exceed feasible deployment levels for each technology when such constraints are supported by available data.

2.2  Data Structure and Sources

This study employs the Switch 2.0 model with a customized electrolyzer module, applied to 47 load zones that correspond to Kenya's counties. Existing generation projects are mapped to individual counties based on their geographic locations. Generation costs and capacity limits are primarily drawn from Kihara et al., which synthesizes data from Kenya's most recent LCPDP.[24] Battery storage costs are instead sourced from the 2024 Annual Technology Baseline published by the U.S. National Renewable Energy Laboratory (NREL).[29] Existing transmission infrastructure is simplified by aggregating lines between counties according to substation locations. The geometric centroid of each county is used to estimate the length of existing and candidate transmission lines. New transmission investments are only permitted between directly adjacent counties. Long distance transmission capacity can be inferred as sequential flows across multiple inter-county links. Improved geospatial data for proposed transmission lines could enhance the accuracy of modeled network expansion.

Temporally, this study models planning periods from 2027 to 2050 in one-year increments. Like most capacity expansion models, Switch assumes perfect foresight regarding future electricity demand. Annual demand projections are based on the Transmission Master Plan 2023–2042, published by the Kenya Electricity Transmission Company Limited (KETRACO).[30] For years beyond the forecast horizon (2043–2050), demand was extrapolated based on the average growth rate observed in the final five years of available data. Annual demand is disaggregated by customer class, including industrial, commercial, residential, street lighting, and flagship projects, using the classification from the 2013 Distribution Master Plan developed by Kenya Power and Lighting Company (KPLC).[31] Demand in each customer class is then allocated to counties based on economic output (for industrial and commercial classes) or population (for the residential class). These county-level annual demands are further converted into hourly profiles using the load shape from Carvallo et al.[23] To reduce computational complexity, the model samples six evenly spaced hours from the 24-hour profile to represent daily demand. Intra-annual variability is not modeled, as Kenya's electricity demand does not exhibit pronounced seasonal patterns. The methodology for selecting representative hours follows the approach used by Carvallo et al.[23]

Wind, solar, and hydroelectric generation vary by time of day and season, requiring models to capture their hourly availability to inform investment and dispatch decisions. In capacity expansion models, time-varying capacity factors are critical for evaluating the reliability and economics of adding intermittent renewable resources. To represent solar availability, we simulated hourly output from a 1 kW photovoltaic system located at the geometric centroid of each county using the National Renewable Energy Laboratory's System Advisor Model (SAM). For wind generation, we applied hourly capacity factor profiles from Carvallo et al. for counties identified in the LCPDP as having viable wind potential.[23] Hydropower in Kenya exhibits seasonal variation; therefore, we used the annual average capacity factors reported by Kihara et al.[24] While this approach enables tractable modeling of spatially distributed renewables, it does not capture potential sub-county variation. In particular, using centroid-based profiles may over- or under-estimate actual generation potential in areas with heterogeneous resource availability. The same set of hourly capacity factors is applied each year across the entire modeling horizon, meaning the model does not account for inter-annual variability in resource availability due to climate or weather fluctuations.

To ensure comparability with the Kenyan strategy, we used hydrogen parameters from its base case assumptions.[9] However, the strategy does not include projections for future cost reductions in electrolyzer technologies. Zun and McLellan evaluated global cost trajectories for green hydrogen under various scenarios, accounting for factors such as economies of scale, learning effects, and research and development.[32] From these, we apply the most conservative cost reduction pathway, which assumes only R&D-driven improvements. In this scenario, electrolyzer capital costs decline by 4.3 percent per year, beginning at an initial cost of $800/kW as stated in the Kenyan strategy. Hydrogen is priced at $2/kg, a widely recognized threshold for cost competitiveness with fossil fuel alternatives (i.e. cost parity).[12,33,34] This price assumption is particularly important, as the model presumes that all hydrogen produced is sold. A competitive price point is therefore necessary to justify this assumption.

2.3  Scenarios

The data assumptions described in section 2.2, with hydrogen data excluded, forms the basic business-as-usual (BAU) scenario, which is used as a baseline to compare the impacts of hydrogen electrolyzers. Based on BAU, we added the hydrogen data with the hydrogen electrolyzer module to compare the impacts of electrolyzers on the grid. In alignment with the Kenyan strategy, we constructed a hydrogen base case (Hydrogen Strategy scenario) where electrolyzer capacity is capped at 100 MW by 2027 and 250 MW by 2032. This is also because hydrogen electrolysis is an emerging technology without extensive proven commercial projects. The practical constraints around adopting new technologies will render unlimited electrolyzer installation in Kenyan infeasible in the near term.

In addition to varying hydrogen-related parameters, we constructed additional scenarios by modifying two sets of input data from the core Switch model that are not directly linked to hydrogen production or sales. Each additional scenario was run both with and without hydrogen electrolyzers to isolate their impacts. First, results from the hydrogen base case reveal a strong correlation between installed wind capacity and electrolyzer deployment. However, Kenya's current Least Cost Power Development Plan (LCPDP) includes only a limited set of new wind projects. To assess how expanded wind deployment might influence the benefits of electrolyzers, we varied the first year when wind projects beyond those identified in the LCPDP are permitted to enter the system, testing start dates from 2028 to 2045.

Second, while Switch assumes perfect foresight of electricity demand, long-term demand projection remains one of the most uncertain aspects of power system planning. In Kenya, electricity demand is shaped by dynamic trends in urbanization, industrial growth, and electrification, all of which carry significant uncertainty over multi-decade horizons. Sensitivity analysis is therefore essential to evaluate whether investment decisions remain robust under divergent growth trajectories. To capture this uncertainty, we constructed low and high demand scenarios using the "Low" and "Vision" cases from KETRACO's transmission master plan.[30] In our model, the Reference scenario from KETRACO corresponds to the BAU case and assumes an average annual growth rate of approximately 5 percent. The Low and Vision scenarios assume average growth rates of roughly 4 percent and 8 percent, respectively. As a result, 2050 electricity demand in the high and low scenarios differs from the reference case by +70 percent and -13 percent, respectively.

## 3 Results and Discussions

### 3.1 Grid Impacts from Hydrogen Electrolyzers

### 3.1.1 National Impacts

Kenya's grid is already largely powered by renewable energy. As of June 2024, renewable sources account for approximately 80 percent of installed capacity and more than 90 percent of domestic electricity generation.[35] Our model projects that this trend will continue, even under the business-as-usual (BAU) scenario (**Figure 1**). Geothermal remains a dominant source due to its suitability as a baseload technology. Over the long term, wind generation expands significantly and is accompanied by corresponding investments in battery storage. The increasing share of wind in Kenya's generation mix is the primary driver of declining levelized cost of electricity (LCOE), underscoring the importance of system flexibility in reducing overall electricity costs.

Although the grid remains predominantly renewable, fossil fuel generation fluctuates between 3 and 13 percent of the total mix through 2042.

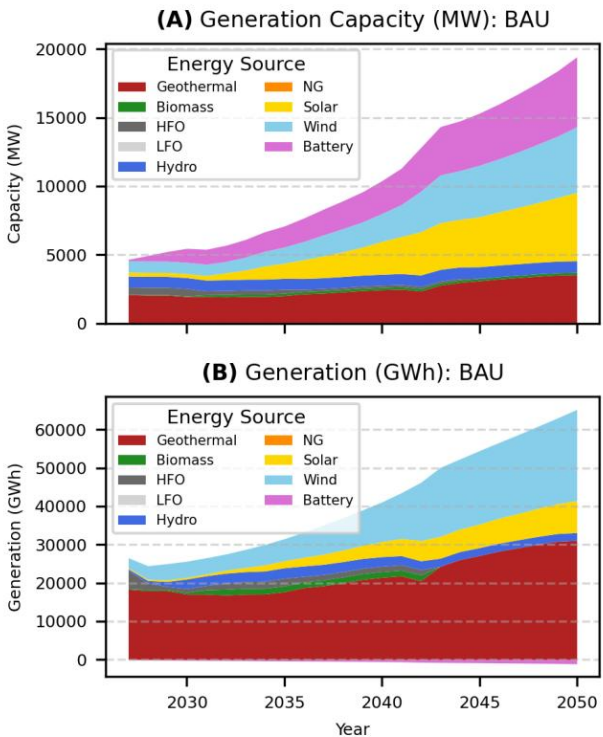

**Figure 1**. Annual generation capacity installed (A) and annual electricity generation (B) for BAU scenario. Battery stores and discharges energy, so only its net production is shown in (B).

As discussed in Section 2.2, in the Hydrogen Strategy scenario, the model is set to sell all hydrogen produced at a fixed price of $2/kg. Before 2030, electrolyzer capacity is capped at levels outlined in Kenya's national strategy—100 MW by 2027 and 250 MW by 2030, generating more than 36,000 tons of hydrogen annually. Beyond 2030, the model allows unrestricted expansion of electrolyzer capacity. Results from the Hydrogen Strategy scenario shows that the model replicates the Kenyan target by building 250 MW of electrolyzers by 2028 and begins expanding capacity further in 2032. By 2050, the system installs approximately 5.5 GW of electrolyzers while achieving total system cost savings of $460 million over the entire planning period. These cost savings begin to accrue immediately upon electrolyzer deployment and grow steadily over time. More than 555,000 tons of hydrogen will be generated annually.

Most of the savings are realized through increased investment in wind generation. As shown in **Figure 2** (C), installed wind capacity rises sharply following the introduction of electrolyzers. This capacity expansion translates into higher overall wind generation, illustrated in **Figure 2** (B) where wind appears in light blue. The operational flexibility provided by electrolyzers also reduces curtailment rate of other baseload renewables such as hydropower or geothermal by 15 to 38 percentage points over the planning period. Because installed hydropower capacity remains constant, this increase in effective utilization lowers the levelized cost of hydropower. At the same time, the model shows reductions in both installed capacity and output for geothermal and solar resources. This is consistent with their higher cost relative to wind in Kenya, which stems from geothermal's high capital costs and solar's lower availability. Together, the reduced energy curtailment, increased use of low-cost wind, and displacement of higher-cost geothermal and solar generation lead to a substantial decline in system-wide levelized cost of electricity (LCOE). LCOE in the Hydrogen Strategy case is approximately 6 percent lower by 2030 and 30 percent lower by 2050 compared to BAU, and is approximately 40 percent lower by 2050 compared to 2027.

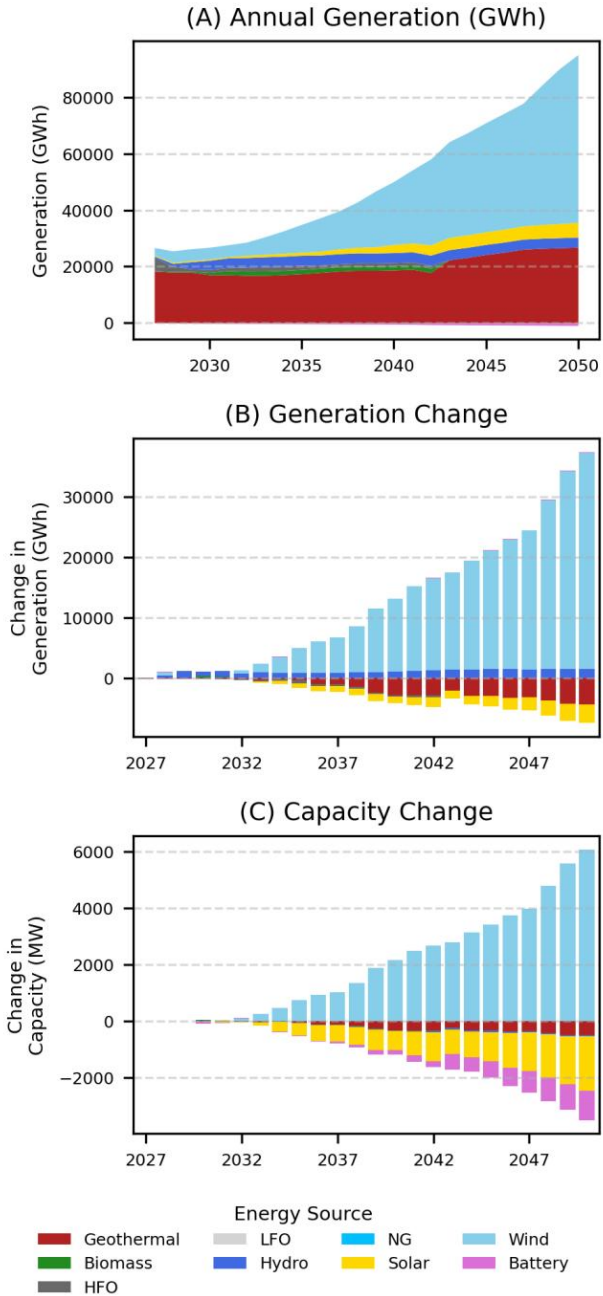

**Figure 2**. Annual electricity generation (GWh) (A), difference in electricity generation (GWh) compared to BAU (B), and difference in installed capacity (MW) compared to BAU (C) in the Hydrogen Strategy scenario.

The integration of grid-connected hydrogen electrolyzers into Kenya's hydrogen strategy induces a fundamental shift not only in the configuration of the power system but also in its operation. Early deployment of electrolyzers is constrained by policy assumptions derived from the national strategy, yet the model shows that even modest capacity contributes meaningfully to system efficiency. Electrolyzers help reduce curtailment of renewable generation, especially from hydropower in the early years (**Figure 3** (A)). Electrolyzers absorb surplus wind energy during periods of moderate demand, such as shoulder hours, thereby reducing hydropower curtailment and improving overall system efficiency. Later in the planning period, as wind capacity expands, electrolyzer utilization increases significantly. By 2050, electrolyzers function as a major source of flexible demand, operating primarily during periods of surplus renewable output, particularly from wind, and drawing up to 6,000 MW during peak dispatch hours (**Figure 3** (B) and (C)). Grid-connected hydrogen serves as a flexible demand sink that better fits the demand curve to the shape of the supply curve. It proves to be an effective demand-side management tool, reducing curtailment and enabling higher penetration of variable renewables.

While **Figure 2** shows a reduction in total installed battery capacity, battery dispatch remains largely stable, as illustrated by the pink lines in **Figure 3**. Across years and scenarios, charge–discharge patterns and magnitude are largely consistent: batteries charge during periods of renewable energy surplus and discharge during evening demand peaks. This suggests higher utilization rate for the remaining storage assets. In fact, the annual energy throughput ratio of batteries increases by 1.5 percentage points on average. Rather than displacing battery storage, electrolyzers operate in parallel, with each technology fulfilling distinct and complementary grid-balancing roles. For instance, during the peak hour of the representative day depicted in **Figure 3**, wind generation declines while electrolyzer output remains steady; this is enabled by battery discharging, which meets demand and supports continued hydrogen production.

In contrast, the integration of electrolyzers reduces – but not eliminates – the role of geothermal and solar generation in later years. Both technologies exhibit a decline in installed capacity, annual output, and operational dispatch. Because geothermal and solar are more expensive than wind in the Kenyan context, this shift underscores the role of electrolyzers in improving overall system cost-efficiency by facilitating greater wind integration. Nonetheless, geothermal and solar remain integral components of the generation mix, continuing to provide dispatchable and location-diversified resources that enhance system resilience. Geothermal remains the main source to meet baseload, operating with virtually no curtailment, supplemented by hydropower, biomass, and minimal heavy fuel oil plants. Solar provides electricity in counties where simply extending and expanding transmission capacity is not cost-optimal to meet existing demand. Almost no county hosts more than one type of renewable energy resource (i.e. wind, solar, geothermal, and hydropower), highlighting the spatial importance of maintaining a meaningful profile of diverse electricity mix.

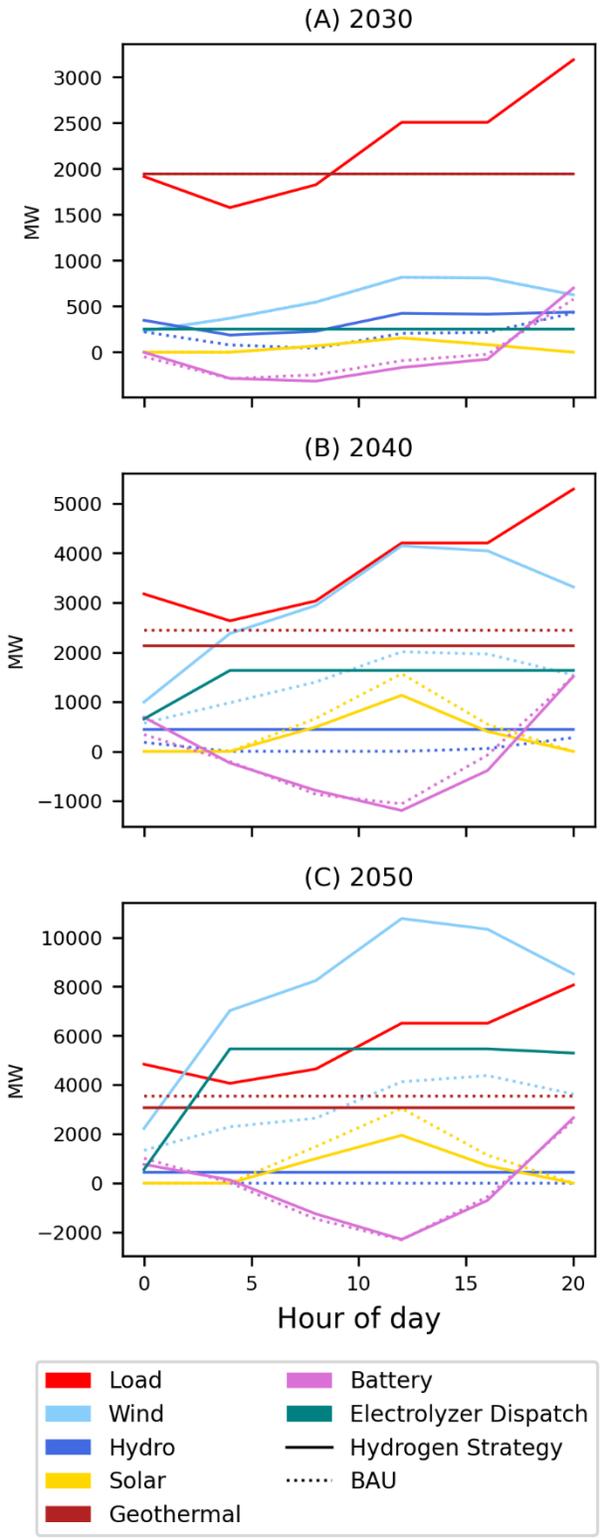

**Figure 3**. National load and dispatch curves of renewable technologies in key years of a representative day. Note that only 6 hours are modeled in a single day.

3.1.2 Subnational Opportunities to Leverage Development and the Just Transition

While national-level indicators such as total installed capacity or system-wide LCOE are commonly used to evaluate energy system transitions, they often obscure important spatial heterogeneity. The introduction of grid-connected electrolyzers fundamentally alters the geography of electricity generation and infrastructure investment in Kenya, leading to uneven distributions of capacity, generation, and transmission at the subnational level. These dynamics have significant implications for both the equity and resilience of the energy transition.

In the Hydrogen Strategy scenario, electrolyzers are predominantly sited in counties with high-quality renewable energy resources and available grid infrastructure. This results in an accelerated buildout of wind capacity in Marsabit and Kajiado, both of which are co-located with electrolyzers to take advantage of temporal alignment between renewable availability and flexible hydrogen demand. By 2050, Marsabit and Kajiado together account for nearly 50 percent and 57 percent of the country's total and additional installed generation capacity, representing a marked spatial concentration of new investment. In contrast, the BAU scenario sees 5 and 8 counties contributing to similar shares of total and additional capacity. The change in distribution of installed capacity reflects changes in electricity generated. **Figure 4** shows that in 2050, the Hydrogen Strategy Scenario observes fewer but bigger exporters near Nairobi shown in deeper purple, the biggest consumer of electricity, and surrounding counties become increasingly dependent on import as shown in darker orange. While the national system appears increasingly interconnected, the economic and infrastructural benefits of the hydrogen transition accrue disproportionately to a small number of counties.

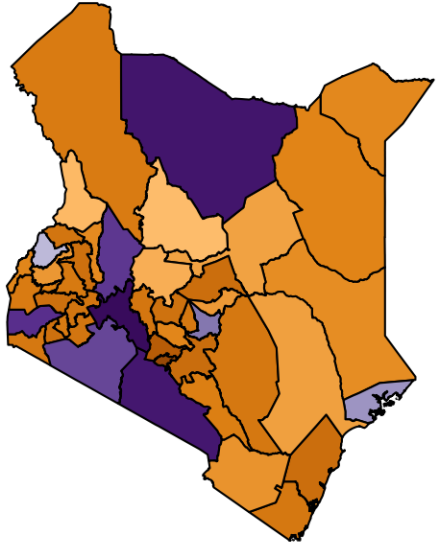

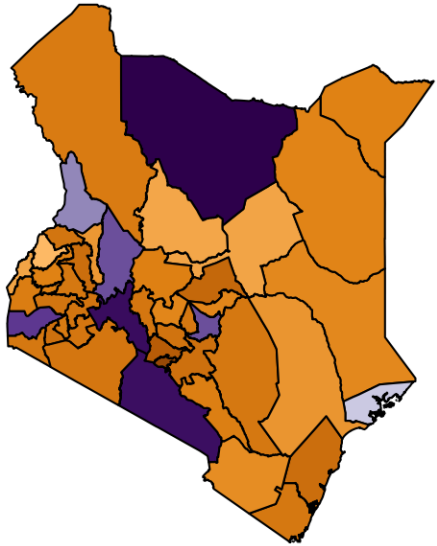

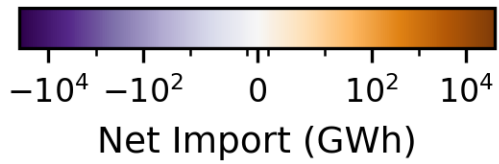

**Figure 4**. Net import and export by county in 2050 for BAU and Hydrogen Strategy Scenarios. Counties shaded in purple are net exporters, while those shaded in orange are net importers.

These transformations are accompanied by a substantial reconfiguration of the transmission network. **Figure 5** shows that in the Hydrogen Strategy scenario, a high-capacity north–south corridor emerges to connect generation nodes in Marsabit and Kajiado – also counties with most electrolyzer capacity – to demand centers in central and southern Kenya. In contrast, counties that see reduced generation under the Hydrogen Strategy scenario, such as Nakuru as well as most western counties, also experience a decline in new transmission investment. This spatial divergence in transmission infrastructure highlights the tight coupling between hydrogen siting, renewable resource exploitation, and grid reinforcement.

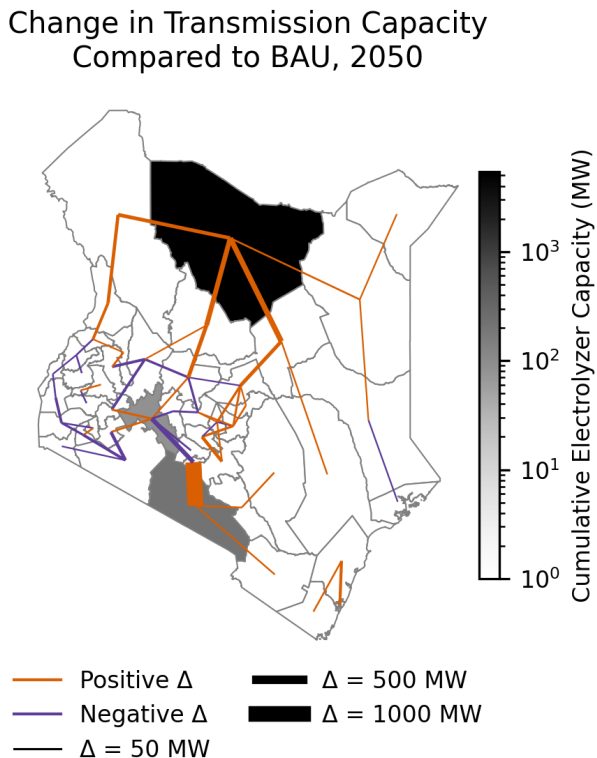

**Figure 5**. Major changes in aggregate transmission capacity between counties in Hydrogen Strategy Scenario compared to BAU in 2050. Changes smaller than 10MW is omitted from the graph for clearer visualization of meaningful changes.

Taken together, these trends raise critical questions about spatial equity in Kenya's energy transition. While the deployment of grid-connected hydrogen improves national-scale metrics – lowering system costs and enabling greater renewable integration, it may also concentrate economic benefits in fewer counties. Counties without strong wind potential or high electricity demand, which is often a sign of strong economic growth, may be marginalized in the emerging hydrogen economy. To mitigate these risks, planners and policymakers should consider incorporating equity-oriented siting criteria, targeted transmission upgrades, and benefit-sharing mechanisms to ensure that the hydrogen transition supports inclusive development across Kenya's diverse counties.

3.2    Viability of the Green Hydrogen Industry

3.2.1    Levelized Cost of Hydrogen

Grid-connected hydrogen improves both the performance of the electricity system and the economic viability of Kenya's green hydrogen industry. The Kenyan strategy identifies high levelized cost of hydrogen (LCOH) as a key barrier to competitiveness in the global hydrogen market. Elevated LCOH also limits the feasibility of a domestically oriented hydrogen industry, placing additional pressure on Kenyan farmers and consumers who already face high fertilizer costs, as well as on a fiscally constrained government that has historically provided fertilizer subsidies. [13,14,36] This high LCOH is primarily driven by electricity prices, which account for an estimated 50 to 70 percent of total production cost.[9] For standalone hydrogen projects, the Kenyan strategy projects LCOH values ranging from $4-13/kg, depending on the technology. The lowest-cost configuration of standalone projects identified in the Kenyan strategy involves coupling electrolyzers with geothermal plants, achieving an LCOH of $4-5/kg.

Our study shows that, using the same cost assumptions as the Kenyan strategy, grid-connected green hydrogen can achieve an LCOH of $4.8/kg in 2027, within the lowest-cost range projected for standalone systems. This cost decreases sharply as more wind capacity is brought online, reaching below $3.2/kg by 2050, even with conservative projections for reduction of capital costs of electrolyzers. While projected capital cost reductions for electrolyzers contribute to this decline, the largest driver is the reduction in levelized cost of electricity (LCOE) achieved through expanded and optimized renewable generation (Section 3.1). This suggests that improving grid operations and expanding wind integration are more impactful levers for lowering hydrogen costs than technology-specific subsidies alone.

It is important to note that LCOH never drops below $2/kg, the parity selling point assumed in this study. This means that while overall system-wide cost savings are achieved, electrolyzer operators are effectively selling at a loss. In practice, operators could choose to sell at LCOH, though this would prevent them from being cost-competitive in international markets. For domestic applications, persistently high LCOH would impose a green premium on Kenyan consumers and the national government. Kenya currently applies a time-of-use (TOU) electricity tariff that halves the regular rate during off-peak hours for eligible customers.[37] Electrolyzer operators may already benefit from this structure, as shoulder hours partially overlap with existing off-peak periods. A more flexible designation of off-peak hours for electrolyzers could further reduce electricity costs, reflecting their contribution to wind integration. Furthermore, Kenya currently lacks any market mechanism that compensates for grid balancing or flexibility services. To fully capitalize on the system benefits provided by hydrogen, additional mechanisms should be introduced to reward grid services offered by electrolyzers. Such policies could also support the viability of battery storage, which provides similar services as illustrated in **Figure 3**.

### 3.2.2 Carbon Intensity and Certification of Grid-Connected Hydrogen

Certification presents another key barrier to the deployment of green hydrogen, particularly for grid-connected systems. Unlike off-grid configurations, grid-connected hydrogen does not draw from a single, traceable energy source. Whether the resulting hydrogen qualifies as green depends on the carbon intensity of the electricity supplied by the grid and how the grid is configured and operated to accommodate the additional load from electrolyzers. Most certification schemes and green hydrogen standards specify a carbon intensity threshold based on the total life-cycle emissions associated with producing 1 kg of hydrogen. Chile has one of the most relaxed requirements at 4.0 kg $CO_2$e/kg $H_2$.[38] The European Union (EU) follows a compliance threshold under Renewable Energy Directive II – Renewable Fuels of Non-Biological Origin (RED II RFNBO) similar to Japan's standard at 3.38 kg $CO_2$e/kg $H_2$ and 3.4 kg $CO_2$e/kg $H_2$, respectively.[39,40] The U.S. Inflation Reduction Act (IRA) enacted the Section 45V – Clean Hydrogen Production Tax Credit that allows tax credit for hydrogen with less than 4 kg $CO_2$e/kg $H_2$, but provides increasing tax benefits in different tiers as intensity reduces to 2.5 kg $CO_2$e/kg $H_2$, 1.5 kg $CO_2$e/kg $H_2$, and 0.45 kg $CO_2$e/kg $H_2$.[41] The most stringent green hydrogen standard is published by the Green Hydrogen Organisation (GH2), set at 1.0 kg $CO_2$e/kg $H_2$.[42]

Although all certification schemes and standards consider the full life-cycle carbon intensity of hydrogen production, including emissions from electricity and water use, the carbon intensity of electricity is the primary driver of total emissions.[43–46] Model results indicate that, when accounting only for grid electricity, hydrogen produced in Kenya has an average carbon intensity of 2.2 kg $CO_2$e/kg $H_2$ in 2028. This value declines to between 1-1.65 kg $CO_2$e/kg $H_2$ through 2035 and falls below 1 kg $CO_2$e/kg $H_2$ thereafter. 2027 marks the only year when hydrogen carbon intensity exceeds all certification schemes as the carbon intensity reaches nearly 6 kg $CO_2$e/kg $H_2$, reflecting limited renewable generation at the start of the planning period before additional wind capacity becomes available.

Because grid-connected electrolyzers in Kenya and other regions do not necessarily operate at 100% capacity factor, certification standards may increasingly incorporate emission factors with higher temporal resolution. Engstam et al. applied dynamic, hourly emission factors and demonstrated that pairing grid-connected electrolyzers with expanded wind capacity reduces both LCOH and marginal carbon abatement costs.[47] Using a similar method, our model applies hourly emission factors based on the grid's dispatch profile, assuming electrolyzers draw electricity only when operated and that emissions reflect real-time generation mixes. Results show that, aside from the first year, grid-connected hydrogen in Kenya meets the EU's RED II RFNBO carbon intensity threshold. As shown in **Figure 6**, most hydrogen produced will meet the GH2 benchmark of 1 kg $CO_2$e/kg $H_2$ starting in 2036. By 2043, all production qualifies for the highest tier of the U.S. IRA Section 45V tax credit.

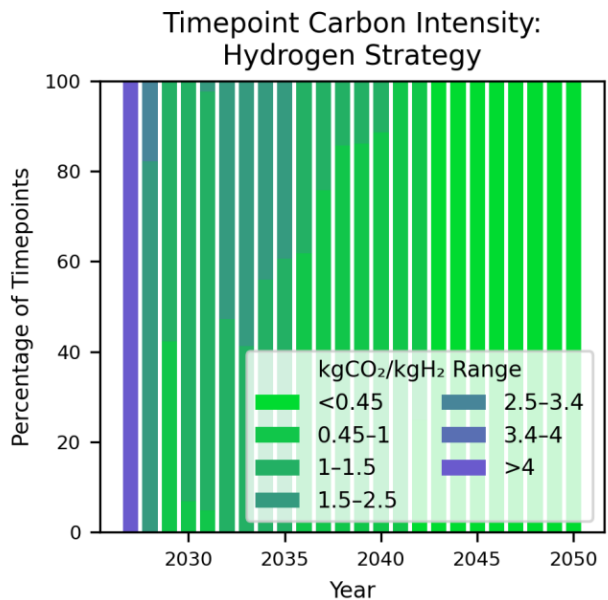

**Figure 6**. Distribution of timepoints by hydrogen carbon intensity in Hydrogen Strategy scenario.

In addition to carbon intensity thresholds, several certification schemes and standards include supplementary criteria, some of which can be assessed through our model. For example, the GH2 standard limits electricity from non-renewable sources to no more than 5 percent.[42] In all but the first two years of the study period, more than 95 percent of Kenya's electricity supply comes from renewable sources. The EU's RED II RFNBO framework further requires compliance with temporal and geographic correlation, as well as additionality, where power purchase agreements (PPAs) can be used as proof.[39,46] While our model does not simulate PPAs between electrolyzers and renewable generators, **Figure 3** demonstrates strong hourly correlation between electrolyzer dispatch and output from wind and hydropower plants. Section 3.1.2 highlighted the co-location of electrolyzers and wind facilities, and Section 3.3 showed alignment between the buildout of new wind projects and electrolyzer capacity. These certification systems also reference broader sustainability principles, including alignment with Environmental, Social, and Governance (ESG) disclosure frameworks and contributions to the Sustainable Development Goals (SDGs). Kenya's green hydrogen industry should remain proactive in understanding and complying with these evolving requirements. Continued engagement by both industry and government in international standard-setting processes will be essential to ensure long-term market access and credibility. Kenya's green hydrogen industry should remain proactive in understanding and complying with these evolving requirements. Continued engagement by both industry and government in international standard-setting processes will be essential to ensure long-term market access and credibility. In addition to aligning with existing frameworks, the Kenyan government can play a more active role in shaping emerging standards by contributing data, sharing implementation experience, and advocating for inclusive and equitable certification criteria that reflect the priorities of developing countries.

## 3.3 Synergistic Planning with Wind Expansion

The benefits of grid-connected hydrogen electrolyzers depend heavily on the structure and timing of renewable energy investment, particularly in wind power. Although electrolyzers can enhance grid flexibility, their full potential for cost savings and emissions reduction is only realized when sufficient renewable capacity is available to meet their demand. In the absence of timely wind expansion, hydrogen deployment is limited, carbon intensity remains elevated, and system-wide benefits are substantially reduced. Kenya's LCPDP includes several utility-scale wind projects totaling up to 300 MW by 2033. However, further investment in wind capacity will be required to fully unlock the economic and environmental advantages of green hydrogen.

**Figure 7** illustrates the impacts of delayed wind expansion on electrolyzer deployment and system costs. Overall, postponing additional wind projects reduces electrolyzer capacity while increasing the system-wide LCOE and total system costs. Even modest delays before 2035 lead to near-term adverse effects. When wind expansion is pushed to 2035 or later, the number of years affected grows substantially, as no new wind capacity is currently planned beyond 2033. Delays extending past 2040 significantly impact the entire planning horizon.

Similar trends are observed for LCOH and hydrogen carbon intensity. Postponing wind additions beyond 2035 results in LCOH remaining flat, with cumulative cost reductions potentially limited to less than 10 percent compared to the model's starting year. Under such scenarios, Kenyan hydrogen may fail to meet the EU's RED II RFNBO carbon intensity threshold for several years, and the timeline for meeting the stricter GH2 standard would also be delayed. These outcomes underscore that delays in wind development not only constrain grid cost savings but also threaten the long-term viability of Kenya's green hydrogen sector.

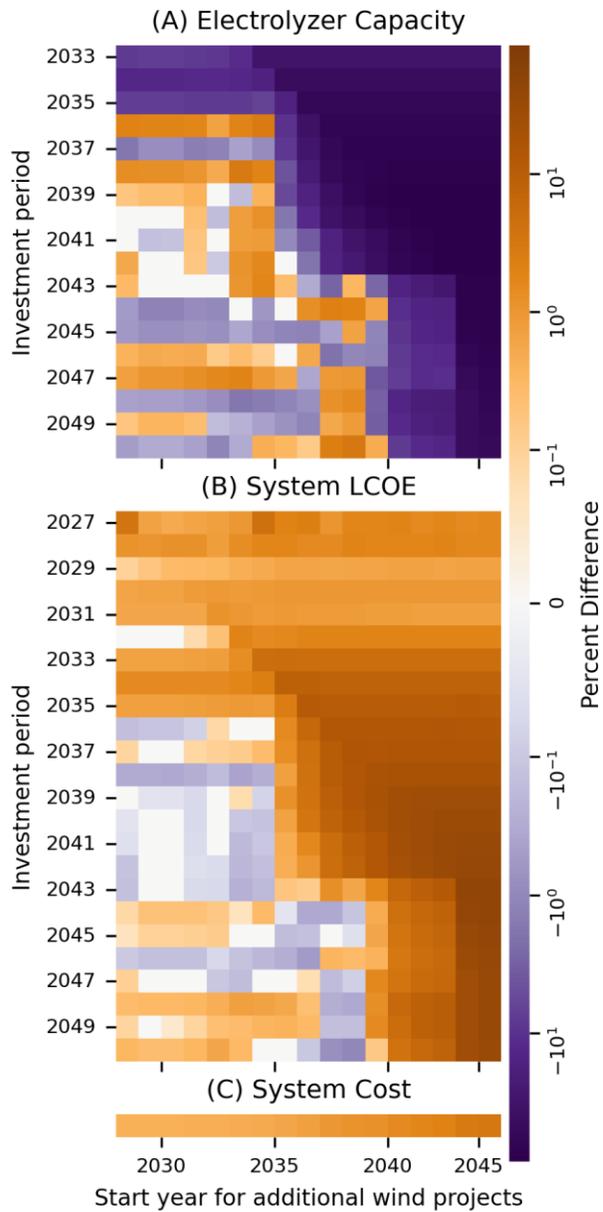

**Figure 7**. Percent change in installed electrolyzer capacity (A), system-wise LCOE (B) and total system cost over the entire planning period (C) when we delay the first year when additional wind projects are allowed to be brought online compared to the Hydrogen Strategy scenario.

These results underscore the importance of co-optimizing hydrogen and wind deployment. If electrolyzer capacity expands without a corresponding increase in wind generation, the electricity used for hydrogen production may be drawn from additional fossil resources or compete with existing loads, undermining both climate and economic goals. Conversely, if wind is developed without adequate flexible demand, curtailment may rise and capacity factors may decline, reducing returns on investment and potentially weakening investor confidence in utility-scale renewables. Planning electrolyzers and wind in tandem ensures that clean electricity enables low-carbon hydrogen, while hydrogen demand enhances renewable integration.

Existing analyses and planning processes often treat hydrogen and renewables as separate tracks, evaluated through parallel policy and procurement frameworks, and risk locking in suboptimal investments. Our study suggests that maximizing the benefits of hydrogen requires not only integrating electrolyzer operation into grid planning, but also coordinating wind investment timelines, transmission development, and the design of flexibility markets. Kenya's green hydrogen strategy will be most effective when implemented as part of a broader, system-wide energy transition that capitalizes on the complementary roles of flexible demand and variable renewable generation.

## 3.4 Sensitivity to Demand Change

Electricity demand growth in Kenya will play a central role in shaping the outcomes of long-term power sector planning. While current projections anticipate rising demand due to electrification, industrial expansion, and regional integration, the pace and magnitude of this growth remain uncertain. This uncertainty has direct implications for investment decisions in generation, transmission, and flexible technologies such as electrolyzers. In this section, we evaluate how variation in electricity demand affects the performance and implications of the Hydrogen Strategy scenario by drawing on different KETRACO demand projection scenarios (as discussed in Section 2.3). Specifically, we assess two questions: first, whether the cost, emissions, and capacity benefits associated with hydrogen deployment hold across low, reference, and high demand trajectories; and second, whether the system-level actions identified in previous sections—particularly wind expansion and transmission reinforcement—remain necessary regardless of the demand scenario. These analyses test the robustness of our findings and help inform planning priorities under conditions of demand uncertainty.

The benefits of electrolyzer integration persist across all demand trajectories, with their magnitude scaling proportionally to electricity demand. In the low and high demand Hydrogen Strategy scenarios (LDH2 and HDH2, respectively), electrolyzers continue to yield system-wide cost savings of $394 million and $829 million by 2050. Electrolyzer capacity expands more aggressively under higher demand: while installed capacity remains similar across scenarios in the first six years, it reaches 4.2 GW in LDH2, 5.5 GW in the Hydrogen Strategy scenario, and 14.2 GW in HDH2 by 2050. This variation reflects increased system flexibility and greater availability of surplus renewable generation under high-demand conditions, which improves the economics of hydrogen production.

System-wide reductions in levelized cost of electricity (LCOE) also remain consistent. In LDH2, LCOE is less than 4 percent higher than in the Hydrogen Strategy reference case but still 33 percent lower than in a low demand case without hydrogen (LD). In HDH2, electrolyzer integration reduces LCOE by an additional 10 percent relative to the Hydrogen Strategy scenario, and by 44 percent compared to a high demand case without hydrogen (HD). These cost advantages are driven primarily by expanded wind deployment. By 2050, wind supplies 59 percent of total electricity generation and accounts for 47 percent of installed capacity in LDH2. In HDH2, these shares rise to 76 percent and 54 percent, respectively. **Figure 8** illustrates this trend: while geothermal, solar, and battery capacity vary across demand scenarios, wind consistently emerges as the primary enabler of hydrogen deployment.

These findings indicate that the case for integrating grid-connected hydrogen electrolyzers remains strong despite uncertainty in electricity demand. Electrolyzer deployment delivers cost savings and contributes to decarbonization even under conservative demand assumptions, and its benefits increase with higher demand. Grid-connected hydrogen also meets international carbon intensity benchmarks across all scenarios. Annual hydrogen carbon intensity remains at least 15 percent below the EU RED II RFNBO threshold, and the year when intensity falls below 1 kg $CO_2$/kg $H_2$ is delayed by only 1–2 years in low demand conditions. LCOH also improves under higher demand, falling an additional 5 percent in HDH2 compared to the reference Hydrogen Strategy by 2050. Together, these results support early, coordinated investment in wind and hydrogen infrastructure, and suggest that policy and planning can move forward despite uncertainty in future electricity demand.

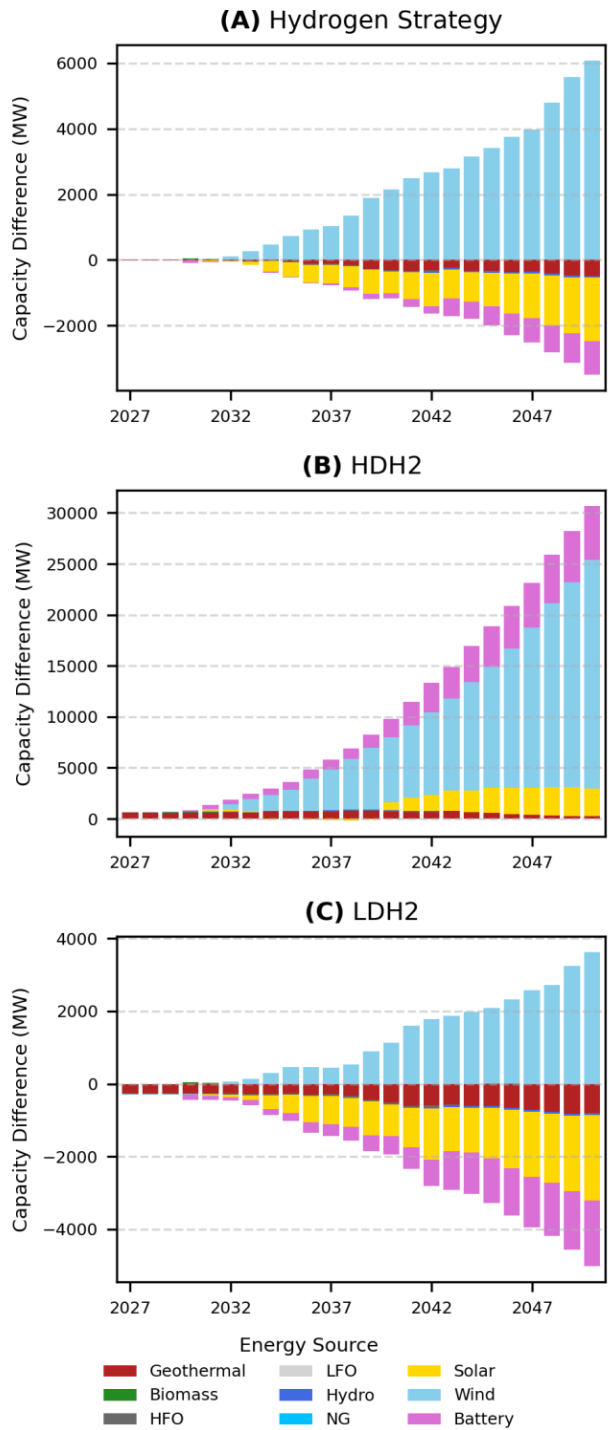

**Figure 8**. Difference in installed capacity by energy source in different demand scenarios when hydrogen electrolyzers are introduced to the grid.

## 3.5 Policy Recommendations

Green hydrogen has the potential to provide a spectrum of benefits to Kenya and the world or to become a new form of colonial energy extractivism. An effective green hydrogen policy should not only ensure that green hydrogen is viable in Kenya but also leverage the industry to provide broader benefits to the country. Grid-connected hydrogen can achieve both by expanding wind capacity and lowering system-wise electricity costs without compromising the emission intensity of green hydrogen. To fully realize the benefits of grid-connected hydrogen, Kenya should integrate hydrogen deployment into its broader power system planning framework. Hydrogen should not be treated as a standalone sector but incorporated into national energy models and infrastructure plans such as LCPDP. Planning electrolyzer capacity in parallel with wind generation, geothermal, solar, and battery storage investments—alongside transmission upgrades—will ensure that hydrogen deployment reinforces, rather than competes with, broader goals of affordability, reliability, and decarbonization. While wind offers the strongest synergy with hydrogen due to its variability and low marginal cost, a diversified mix of renewables and storage technologies contributes to a more flexible and resilient grid, capable of supporting both hydrogen production and wider system needs. The planning process should also proactively consider subnational impacts of infrastructure development to ensure energy security and that costs and benefits associated with development are distributed justly across counties.

Synergistic policies are also important for improving the economics of hydrogen electrolyzers by not only providing lower electricity costs but also additional compensation for the flexibility service that electrolyzers provide. Kenya's time-of-use (TOU) tariff could be refined to better accommodate the operating profiles of electrolyzers, allowing them to respond to low-cost, high-renewable periods. Introducing market mechanisms to reward grid-balancing and flexibility services would also improve the economics of hydrogen and battery storage, ensuring long-term financial viability of projects that enable a renewable-dominant grid in the long-term.

Finally, Kenya should play a more active role in shaping international hydrogen certification standards. As one of the few African countries with a domestic-use strategy, Kenya brings a unique perspective to global forums. By contributing empirical data, sharing implementation experience, and advocating for equitable criteria that reflect the realities of developing countries, Kenya can help shape a just green hydrogen economy free of energy extractivism and colonialism endemic in Africa and other previously colonized states.

## ASSOCIATED CONTENT

The model and dataset associated with this paper is available at: https://github.com/NotEleven/Switch-Kenya-2025_public

## ACKNOWLEDGMENT

We gratefully acknowledge Martin Mutembei and Anne Nganga (Strathmore Energy Research Center), as well as Steve Pye and Pietro Lubello (University College London), for their thoughtful feedback and support throughout this study. We also thank Aashika Nair, Sanya Kwatra, Risper Rwengo, and Theo Xixun Wang for their contributions to data collection and model preparation.